\documentstyle[aps]{revtex}
\begin{document}
\draft

\title{{NSF-ITP-01-18}\\
\ \ \\
\ \ \\
\ \ \\
\ \ \\
Compositional Representation of Protein Sequences\\
and the Number of Eulerian Loops}
\author{Bailin Hao\footnote{Corresponding author. E-mail: hao@itp.ac.cn}
\footnote{On leave from the Institute of Theoretical Physics,
Academia Sinica, P. O. Box 2735, Beijing 100080, China}}
\address{Institute for Theoretical Physics, UCSB, Santa Barbara, CA 93106-4030, USA}
\author{Huimin Xie}
\address{Department of Mathematics, Suzhou University, Suzhou 215006, China}
\author{Shuyu Zhang}
\address{Institute of Physics, Academia Sinica, P. O. Box 603, Beijing 100080, China}
\date{\today}
\maketitle
\begin{abstract}
An amino acid sequence of a protein may be decomposed into consecutive overlapping
strings of length $K$. How unique is the converse, i.e., reconstruction of amino acid
sequences using the set of $K$-strings obtained in the decomposition? This problem may be
transformed into the problem of counting the number of Eulerian loops in an Euler graph,
though the well-known formula must be modified. By exhaustive enumeration and by using
the modified formula, we show that the reconstruction is unique at $K\geq 5$ for an
overwhelming majority of the proteins in {\sc pdb.seq} database. The corresponding Euler
graphs provide a means to study the structure of repeated segments in protein sequences.
\end{abstract}

\section{Introduction}

The composition of nucleotides in DNA sequences and the amino acids composition
in protein sequences have been widely studied. For example, the $g+c$ contents or
$CpG$ islands in DNAs have played an important role in gene-finding programs.
However, this kind of study usually has been restricted to the frequency of single
letters or short strings, e.g., dinucleotide correlations in DNA sequences\cite{wli},
amino acids frequency in various complete genomes\cite{proteome}. However, in contrast to DNA
sequences amino acid correlations in proteins have been much less studied. A simple reason
might be that there are 20 amino acids and it is difficult to comprehend the 400
correlation functions even at the two-letter level. A more serious obstacle consists in
that protein sequences are too short for taking averages in the usual definition of
correlation functions.  

For short sequences like proteins one should naturally approach the problem from
the other extreme by applying more deterministic, non-probabilistic methods. In fact,
the presence of repeated segments in a protein is a strong manifestation of amino acid
correlation. This problem has a nice connection to the number of Eulerian loops in
Euler graphs. Therefore, we start with a brief detour to graph theory. 
 
\section{Number of Eulerian Loops in an Euler Graph}

Eulerian paths and Euler graphs comprise a well-developed chapter of graph
theory, see, e.g., \cite{fleischner}. We collect a few definitions in order to fix our
notation. Consider a connected, directed graph made of a certain number of labeled nodes.
A node $i$ may be connected to a node $j$ by a directed arc. If from a starting node $v_0$
one may go through a collection of arcs to reach an ending node $v_f$ in such a way that
each arc is passed once and only once, then it is called an {\it Eulerian path}. If $v_0$
and $v_f$ coincide the path becomes an {\it Eulerian loop}. A graph in which there exists
an Eulerian loop is called an {\it Eulerian graph}. An Eulerian path may be made an
Eulerian loop by drawing an auxiliary arc from $v_f$ back to $v_0$. We only consider Euler
graphs defined by an Eulerian loop.

From a node there may be $d_{\rm out}$ arcs going out to other nodes, $d_{\rm out}$ is
called the outdegree (fan-out) of the node. There may be $d_{\rm in}$ arcs coming into
a node, $d_{\rm in}$ being the indegree (fan-in) of the node. The condition for a graph
to be Eulerian was indicated by Euler in 1736 and consists in
 $$d_{\rm in}(i)=d_{\rm out}(i)\equiv d_i ={\rm an\,\,\,even\,\,\,number}$$ 
for all nodes~$i$.

Numbering the nodes in a certain way, we may put their indegrees as a diagonal matrix:
\begin{equation}
\label{mmatrix}
M={\rm diag}(d_1, d_2, \cdots, d_m).
\end{equation}

The connectivity of the nodes may be described by an adjacent matrix $ A=\{a_{ij}\}$,
where $a_{ij}$ is the number of arcs leading from node $i$ to node $j$.

From the $M$ and $A$ matrices one forms the Kirchhoff matrix:
\begin{equation}
\label{cmatrix}
C=M-A.
\end{equation}

The Kirchhoff matrix has the peculiar property that its elements along
any row or column sum to zero: $\sum_i c_{ij}=0$, $\sum_j c_{ij}=0$.
Further more, for an $m\times m$ Kirchhoff matrix all $(m-1)\times(m-1)$ minors
are equal and we denote it by $\Delta$.

A graph is called simple if between any pairs of nodes there are no parallel
(repeated) arcs and at all nodes there are no rings, i.e., $a_{ij}=0$ or
$1 \,\,\,\forall i, j$ and $a_{ii}=0\,\,\,\forall i$. The number $R$ of Eulerian
loops in a simple Euler graph is given by\\
{\bf The BEST Theorem}\cite{fleischner} (BEST stands for N. G. de {\bf B}ruijn,
 T. van Aardenne-{\bf E}hrenfest, C. A. B. {\bf S}mith, and W. T. {\bf T}utte):
\begin{equation}
\label{best}
R =\Delta \prod_i (d_i-1)!
\end{equation}

For general Euler graphs, however, there may be arcs going out and coming into
one and the same node (some $a_{ii}\neq 0$) as well as parallel arcs leading from
node~$i$ to~$j$ ($a_{ij}> 1$). It is enough to put auxiliary nodes on each parallel
arc and ring to make the graph simple. The derivation goes just as for simple
graphs and the final result is one has the original graph without auxiliary
nodes but with $a_{ii}\neq 0$ and $a_{ij}> 1$ incorporated into the adjacent
matrix $A$. However, in accordance with the unlabeled nature of the parallel
arcs and rings one must eliminate the redundancy in the counting result by
dividing it by $a_{ij}!$. Thus the BEST formula is modified to 
\begin{equation}
\label{bestxie}
R =\displaystyle\frac{\Delta \prod_i (d_i-1)!}{\prod_{ij}a_{ij}!}
\end{equation}
As $0!=1!=1$ Eq.~(\ref{bestxie}) reduces to (\ref{best}) for simple graphs.

\section{Eulerian Graph from a Protein Sequence}

We first decompose a given protein sequence of length $L$ into a  
set of $L-K+1$ consecutive overlapping $K$-strings by using a
window of width $K$, sliding one letter at a time. Combining repeated
strings into one and recording their copy number, we get a collection
$\{W^K_j, n_j\}^M_{j=1}$, where $M\leq L-K+1$ is the number of different
$K$-strings. 

Now we formulate the inverse problem. Given the collection $\{W^K_j, n_j\}^M_{j=1}$
obtained from the decomposition of a given protein, reconstruct all possible
amino acid sequences subject to the following requirements:
\begin{enumerate}\itemsep 0pt
\item Keep the starting $K$-string unchanged. This is because most protein sequences
start with methionine ({\tt M}); even the tRNA for this initiation~{\tt M}
is different from that for elongation. This condition can easily be relaxed.
\item Use each $W^K_j$ string $n_j$ times and only $n_j$ times until the given collection
is used up.
\item The reconstructed sequence must reach the original length~$L$.
\end{enumerate}
Clearly, the inverse problem has at least one solution --- the original protein
sequence. It may have multiple solutions. However, for $K$ big enough the solution
must be unique as evidenced by the extreme case $K=L-1$. We are concerned with
how unique is the solution for real proteins. Our guess is for most proteins the
solution is unique at $K\geq 5$.

 In order to tell the number of reconstructed
sequences we transform the original protein sequence into an Euler graph
in the following way. Consider the two $(K-1)$substrings of a $K$-string as two
nodes and draw a directed arc to connect them. The same repeated $(K-1)$-strings
are treated as a single node with more than one incoming and outgoing arcs.

Take the SWISS-PROT entry ANPA\_PSEAM as an example\cite{pdbseq}. This antifreeze
protein A/B precursor of winter flounder has a short sequence of 82 amino acids
and some repeated segments related to alanine-rich helices. Its sequence reads:
\begin{verbatim}
     MALSLFTVGQ LIFLFWTMRI TEASPDPAAK AAPAAAAAPA AAAPDTASDA AAAAALTAAN
     AKAAAELTAA NAAAAAAATA RG
\end{verbatim}
Consider the case $K=5$.
The first 5-string {\tt MALSL} gives rise to a transition from node \fbox{\tt MALS}
to \fbox{\tt ALSL}. Shifting by one letter, from the next 5-string {\tt ALSLF} we get 
an arc from node \fbox{\tt ALSL} to node \fbox{\tt LSLF}, and so on, and so forth.
Clearly, we get an Eulerian path whose all nodes have even indegree (outdegree)
except for the first and the last nodes. Then we draw an auxiliary arc from the
last node \fbox{\tt TARG} back to the first \fbox{\tt MALS} to get a closed
Eulerian loop. 

In order to get the number of Eulerian loops there is no need to generate a fully-fledged
graph with all the $M$ distinct $(K-1)$-strings treated as nodes. The number of nodes may be
reduced by replacing a series of consecutive nodes with $d_{\rm in}=d_{\rm out}=1$ by a
single arc, keeping the topology of the graph unchanged. In other words, only those
strings in $\{W^{K-1}_j, n_j\}$ with $n_j\geq 2$ are used in drawing the graph.
In our example it reduces to a small Euler graph consisting of 9 nodes:
$$\{{\tt AKAA}, 2;\, {\tt AAPA}, 2;\, {\tt APAA}, 2;\, {\tt PAAA}, 2;\, {\tt AAAA}, 10; 
\, {\tt AAAP}, 2;\,  {\tt LTAA}, 2;\,  {\tt TAAN}, 2;\, {\tt AANA}, 2\}.$$
The Kirchhoff matrix is:
\begin{equation}\label{ak}
\begin{array}{lr}
C=\left(\begin{array}{ccccccccc}
2 &-1 & 0 & 0 & 0 & 0 &-1 & 0 & 0\\
0 & 2 &-2 & 0 & 0 & 0 & 0 & 0 & 0\\
0 & 0 & 2 &-2 & 0 & 0 & 0 & 0 & 0\\
0 & 0 & 0 & 2 &-2 & 0 & 0 & 0 & 0\\
-1& 0 & 0 & 0 & 4 &-2 &-1 & 0 & 0\\
0 &-1 & 0 & 0 &-1 & 2 & 0 & 0 & 0\\
0 & 0 & 0 & 0 & 0 & 0 & 2 &-2 & 0\\
0 & 0 & 0 & 0 & 0 & 0 & 0 & 2 &-2\\
-1& 0 & 0 & 0 &-1 & 0 & 0 & 0 & 2\\
\end{array}\right),\\
\end{array}
\end{equation}
The minor $\Delta=192$ and 
$$ R(5)=\frac{\Delta 9!}{6! 2^6}=1512.$$
We write $R(K)$ to denote the number of reconstructed sequences from a decomposition
using $K$-strings.

We note, however, precautions must be taken with spurious repeated arcs caused by the
reduction of number of nodes. In calculating
the $\prod_{ij}a_{ij}$ in the denominator of Eq.~(\ref{bestxie}) one must subtract the
number of spurious repeated arcs from the corresponding matrix element of the adjacent
matrix. This remark applies also to the auxiliary arc obtained by connecting the last
node to the first. Fortunately, there are no such spurious arcs in the example above.

We have written a program to exhaustively enumerate the number of reconstructed
amino acid sequences from a given protein sequence and another program to implement the
Eq.~(\ref{bestxie}). The two programs yield identical results whenever comparable ---
the enumeration program skips the sequence when the number of reconstructed sequences
exceeds 10000. 

\section{Result of Database Inspection}
\label{s4}

We used the two programs to inspect the 2820 proteins in the special selection
{\sc pdb.seq}\cite{pdbseq}. The summary is given in Table~\ref{tab1}. As expected most
of the proteins lead to unique reconstruction even at $K=5$. At $K=10$ such proteins
make 99\% of the total.
\begin{table}
\caption{Distribution of the 2820 proteins in {\sc pdb.seq} by the number of
reconstructed sequences at different $K$. Percentages in parentheses are given
in respect to the total number 2820.}
\label{tab1}
\begin{center}
\begin{tabular}{c|cccccc}
$K$ & Unique &2-10 & 11-100 & 101-1000 & 1001-10000 & $>$ 10000\\
\hline
5  & 2164 (76.7\%)   & 404 & 90     & 45       & 21          & 93      \\
6  & 2651 (94.0\%)  & 77  & 29     & 10       & 4           & 49      \\
7  & 2732 (96.9\%)  &  32 &  16    &  3       &  2          & 44      \\
8  & 2740 (97.1\%)  &  23 & 10     &  3       &  0          & 44      \\
9  & 2763 (97.9\%)  &  13 &  7     &  1       &  0          & 36      \\
10 & 2793 (99.0\%)  &  11 &  7     &  2       &  1          &  6      \\
11 & 2798 (99.2\%)  &  12 &  2     &  1       &  1          &  6      \\
\end{tabular}
\end{center}
\end{table}

The fact that most of the protein sequences have unique reconstruction is not
surprising if we note that for a random amino acid sequence of the
length of a typical protein one would expect $R=1$ at $K=5$, as it is very unlikely that
its decomposition may yield repeated pairs of $K$-strings among the ${20}^5=3 200 000$
possible strings. A more positive implication of
this uniqueness is one may take the collection of $\{W^K_j\}^L_{j=1}$ as an equivalent
representation of the original protein sequence. This may be used in inferring
phylogenetic relations based on complete genomes when it is impossible to start with
sequence alignment. We will report our on-going work along this line in a separate
publication\cite{wanghao}.

A more interesting result of the database screening  consists in there exists a small
group of proteins which have an extremely large number of reconstructed sequences.
The number $R$ is not necessarily related to the length of the protein.
As a rule, long protein sequences, say, with 2000 or more amino acids, tend to have larger
$R$ at $K=5$ or so, but the number drops down quickly. In fact, all 29 proteins in
{\sc pdb.seq} with more than 2115 amino acids have unique or a small number of reconstructed
sequences. Some not very long proteins have much more reconstructions than the long ones.
We show a few "mild" examples in Table~\ref{tab2}.

\begin{table}
\caption{A few examples of protein decomposition with comparatively large $R$ at
$K=5$. AA is the number of amino acids in the protein.}
\label{tab2}
\begin{center}
\begin{tabular}{ccccc}
Protein & MCMI\_YEAST & PLMN\_HUMAN & CENB\_HUMAN & CERU\_HUMAN \\
\hline
AA  & 286   & 810 & 599 & 1065  \\
R(5)  & 7441920   & 3024000 & 491166720  & 3507840   \\
R(6)  & 39312   & 384 & 17421 &  512   \\
R(7)  & 1620   & 192 &  90 & 21   \\
R(8)  & 252   & 96 & 12 & 6  \\
R(9) & 16   & 5 & 4 &  1 \\
R(10)& 2  & 1 & 1 & {}\\
R(11)& 1  & {}& {}\\
\end{tabular}
\end{center}
\end{table}
The inspection is being extended to all available protein sequences in public databases.

\section{Discussion}

In this paper we have given some precise construction and numbers associated with
real protein sequences. Their biological implications have to be yet explored.

As mentioned in Section~\ref{s4}, we have been using the uniqueness of the 
reconstruction for most protein sequences to justify the compositional distance approach
to infer phylogenetic relations among procaryotes based on their complete genomes\cite{wanghao}.
Most of the phylogenetic studies so far consider mutations at the sequence level.
Sequences of more or less the same length are aligned and distances among species 
are derived from the alignments. However, mutations from a common ancestral sequence
reflect only one way of evolution. There might be another way of protein evolution ---
short polypeptides may fuse to form longer proteins. Perhaps our approach may better capture
the latter situation.

The decomposition and reconstruction described in this paper provide a way to
study polypeptide repeats and amino acid correlations. 
The reconstruction problem naturally singles out a small group of proteins that
have a complicated structure of repeated segments. One may introduce further
coarse-graining by reducing the cardinality of the amino acid alphabet according
to their biological properties. This makes the approach closer to real proteins.
Investigation along these lines are under way.

We note that the Eulerian path problem has been invoked in the study of sequencing
by hybridization, i.e., in the context of RNA or DNA sequences, see \cite{pevzner}
and references therein. To the best of our knowledge the modification of the
BEST formula to take into account parallel arcs and rings has not been discussed so
far.

\section*{Acknowledgments}
This work was accomplished during the Program on Statistical Physics and Biological
Information at ITP, UCSB, supported in part by the National Science Foundation under
Grant No. PHY99-07949. It was also supported partly by the Special Funds for Major
State Basic Research Project of China and the Major Innovation Research Project "248"
of Beijing Municipality. BLH thanks Prof. Ming Li for calling his attention to \cite{pevzner}.


\begin{references}
\bibitem{wli}W. Li, The study of correlation structures in DNA sequences --- a critical review,
{\it Computer \& Chemistry} {\bf 21} (1997) 257-172.
\bibitem{proteome}See, for example, the Proteome page of EBI:\\
{\tt http://www.ebi.ac.uk/proteome/}
\bibitem{fleischner} H. Fleischner, {\it Eulerian Graphs and Related Topics},
Part~1, vol.~2, p.~IX80, Elsevier, 1991.
\bibitem{pdbseq} {\sc pdb.seq} is a collection of SWISS-PROT entries that have one or
more pointers to the PDB structural database. In the file associated with SWISS-PROT
Rel. 39 (May 2000) there are 2821 entries. In our calculation we excluded a protein with too
many {\tt X}s (undetermined amino acids). We fetched the file from:\\
{\tt ftp://ftp.cbi.pku.edu.cn/pub/database/swissprot/special\_selections/pdb.seq}
\bibitem{wanghao} Bin Wang, and Bailin Hao, Procaryote phylogeny based on complete genomes
(in preparation).
\bibitem{pevzner} P. Pevzner, {\it Computational Molecular Biology. An Algorithmic Approach},
\SS 5.4, MIT Press, 2000.
\end{references}
\end{document}